\documentclass[conference]{IEEEtran}
\IEEEoverridecommandlockouts

\usepackage{amsmath,amssymb,amsfonts}

\usepackage{cite}
\usepackage{graphicx}
\usepackage{textcomp}
\usepackage{xcolor}
\usepackage{algpseudocode}
\usepackage{threeparttable}
\usepackage{graphics}
\usepackage{epsfig}
\usepackage{textcomp}
\usepackage{xcolor}
\usepackage{pifont}
\usepackage{amssymb}
\usepackage{booktabs}
\usepackage{multirow}
\usepackage{makecell}

\usepackage{hyperref}
\usepackage{xspace}
\usepackage{url}
\usepackage{mathtools}
\newtheorem{definition}{Definition}
\usepackage{algorithm}

\def\BibTeX{{\rm B\kern-.05em{\sc i\kern-.025em b}\kern-.08em
    T\kern-.1667em\lower.7ex\hbox{E}\kern-.125emX}}

\newcommand{\projName}{SnipSnap\xspace}

\definecolor{SpdClr}{HTML}{006600}

\begin{document}



\title{SnipSnap: A Joint Compression Format and Dataflow Co-Optimization Framework\\for Efficient Sparse LLM Accelerator Design}

\author{
     Junyi Wu$^{1,2}$, Chao Fang$^{1,3,\dagger}$, Zhongfeng Wang$^{1,4,\dagger}$ \\
	\IEEEauthorblockA{
            $^1$Nanjing University~$^2$Hong Kong University of Science and Technology~$^3$Shanghai Qi Zhi Institute~$^4$Sun Yat-sen University \\
        Email:
		\{211870122, fantasysee\}@smail.nju.edu.cn, zfwang@nju.edu.cn
    }
}



\maketitle
\renewcommand{\thefootnote}{}
\footnotetext{$^\dagger$Corresponding author. This work was supported by the National Key R\&D Program of China under Grant 2022YFB4400600.}

\begin{abstract}

The growing scale of large language models (LLMs) has intensified demands on computation and memory, making efficient inference a key challenge. While sparsity can reduce these costs, existing design space exploration (DSE) frameworks often overlook compression formats, a key factor for leveraging sparsity on accelerators. This paper proposes SnipSnap, a joint compression format and dataflow co-optimization framework for efficient sparse LLM accelerator design. SnipSnap introduces: (1) a hierarchical compression format encoding to expand the design space; (2) an adaptive compression engine for selecting formats under diverse sparsity; and (3) a progressive co-search workflow that jointly optimizes dataflow and compression formats. SnipSnap achieves 18.24\% average memory energy savings via format optimization, along with 2248.3$\times$ and 21.0$\times$ speedups over Sparseloop and DiMO-Sparse frameworks, respectively.
 
\end{abstract}


\section{Introduction} \label{sec:intro}

The rapid growth of large language models (LLMs) such as LLaMA~\cite{arxiv_2023_llama2} and OPT~\cite{arxiv_2022_opt} has enabled strong performance across diverse tasks~\cite{github-copilot, arxiv_2023_llama2}. However, their excessive model size leads to high memory and computation demands, posing challenges for efficient hardware deployment.

Leveraging sparsity in LLMs emerges as a promising solution. The presence of substantial zero values in weights or activations enables compression~\cite{iclr_2024_relu, nips_2024_sparsellm}, which significantly reduces both memory and computation requirements.
Nevertheless, general processors often struggle with sparse tensor computing due to compression format complexities and irregular memory access patterns. This gap has driven significant focus on sparse accelerators~\cite{jssc_2017_eyeriss, jetcas_2019_eyerissv2, fang2022algorithm, isca_2021_dstc, micro_2019_sparten, fang2022efficient, shi2024bitwave, liu202316, tu2022sdp, fang2024efficient, micro_2018_cambricon-s, micro_2020_procrustes, isscc_2025_hybrid}, whose effectiveness depends heavily on dataflow and compression format design.
Hence, there is a compelling need for a modeling framework to systematically explore these dataflow and compression format design choices, providing early-stage design guidance for developing more efficient sparse LLM accelerators.

While frameworks like Timeloop~\cite{ispass_2019_timeloop}, ZigZag~\cite{tc_2021_zigzag} and LLMCompass~\cite{isca_2024_llmcompass} excel in dataflow optimization, to the best of our knowledge, few~\cite{iiswc_2021_stonne, micro_2023_teaal} support sparse accelerator modeling, while only DiMO-Sparse~\cite{song2024dimo} and Sparseloop~\cite{micro_2022_sparseloop} further support design space exploration. However, overlooking compression format optimization in these sparsity-aware frameworks, hinders sparse accelerators from fully exploiting sparsity benefits for efficient LLM inference. To be specific, current frameworks have three critical challenges on compression format optimization: space representation, format selection, and exploration efficiency.

\textbf{Challenge 1 (Space Representation):} 
Existing sparsity-aware DSE frameworks~\cite{micro_2022_sparseloop, iiswc_2021_stonne, micro_2023_teaal, song2024dimo} lack a general representation of compression formats, which limits their ability to explore the full design space. These frameworks typically assume fixed, user-defined formats, missing opportunities for further optimization. However, compression formats strongly influence energy and performance, as they determine data size and movement patterns, particularly between accelerators and off-chip memory~\cite{isscm_2025_sparse}.
A unified and flexible format representation is essential to unlock the full potential of sparse LLM acceleration.

\textbf{Challenge 2 (Format Selection):}
Choosing the right compression format for LLMs is difficult due to varying sparsity across modules and applications~\cite{iclr_2024_relu, nips_2024_sparsellm}. Sparsity levels can differ significantly across a model~\cite{iclr_2024_relu} and different application scenarios may tolerate different levels of sparsity, while hardware often supports only a single format. An effective selection should consider diverse sparsity patterns and hardware constraints, as most accelerators in practice support only one format to reduce design overhead and complexity.


\textbf{Challenge 3 (Exploration Efficiency):} 
The joint optimization of dataflow and compression formats results in a vast and complex design space. Existing frameworks either rely on slow cycle-level simulations~\cite{iiswc_2021_stonne} or inefficient search methods~\cite{micro_2022_sparseloop}, both of which hinder timely exploration. The added complexity from format optimization worsens this issue. Efficient DSE requires balance accuracy and speed to identify strong designs without exhaustive search.

To address these challenges, we propose \projName, a framework that jointly optimizes compression formats and dataflows for efficient sparse LLM accelerator design. \projName\ introduces a hierarchical encoding of compression formats, an adaptive format selection engine, and a co-search workflow that jointly optimizes compression and dataflow. Unlike prior arts~\cite{iccad_2019_magnet, ispass_2019_timeloop, tecs_2019_dmazerunner, isca_2024_llmcompass, tc_2021_zigzag, iiswc_2021_stonne, micro_2023_teaal, song2024dimo, micro_2022_sparseloop, geens2024energy}, we recognize compression formats as a critical design parameter that impacts memory accesses and storage requirements. The main contributions of \projName~are as follows:

\textbf{For Challenge 1:} We propose a \textbf{hierarchical compression format encoding method} that extends the exploration space for compression. It divides the format encoding space into two parts: the compression pattern subspace, which defines levels and compression operations for each level, and the dimension allocation subspace, which assigns specific dimension sizes to levels. This enhances the expressiveness and flexibility of sparse accelerator design. (Sec.~\ref{subsec:compre_format})

\textbf{For Challenge 2:} We develop an \textbf{adaptive compression engine} that selects the optimal format for tensors with varying sparsity levels. It incorporates a complexity-based penalization rule to prevent over-specialization, an efficiency-oriented dimension allocation strategy that minimizes the cost of compression and decompression, and an importance-based scoring mechanism that weights LLMs according to their significance. Our method reduces memory energy by 18.24\% on average over widely-used formats.(Sec.~\ref{subsec:search_engine})

\textbf{For Challenge 3:} We present a \textbf{progressive co-search workflow} that integrates dataflow and compression format optimization to accelerate the exploration process. It leverages upfront estimation of computation reduction and compression-aware loop allocation to avoid redundant search iterations.
Experimental results show speedups of 2248.3$\times$ over Sparseloop and 21.0$\times$ over DiMO-Sparse.
 (Sec.~\ref{subsec:co-consideration})


\section{Background and Motivation}
\label{sec:bkg}

\subsection{Sparsity in LLMs}

\begin{figure}[t]
    \centering
    \includegraphics[width=0.48\textwidth]{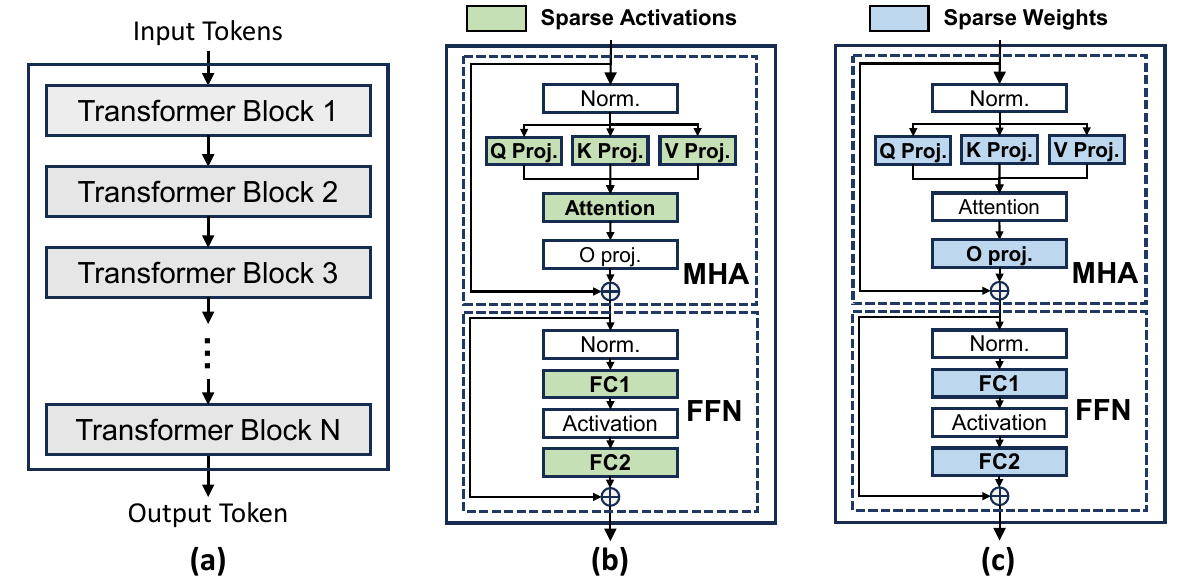}
    \caption{(a) The general architecture of LLMs, while sparsity can be introduced in (b) activation or (c) weight tensors.}
    \label{fig:llm_bkg}
    \vspace{-0.5cm}
\end{figure}

As shown in Fig.~\ref{fig:llm_bkg}, LLMs are stacked by multiple transformer blocks, each containing a multi-head attention (MHA) layer and a feedforward neural network (FFN) layer. The MHA layer mainly includes projections for query ($Q$), key ($K$), value ($V$), and output ($O$), while an FFN layer mainly consists of up-projection ($FC1$) and down-projection ($FC2$) layers. As model size grows, the associated memory and computation demands pose significant challenges for efficient inference.

Sparsity offers a promising path to efficiency by avoiding computation and memory access for zero values. Extensive research has explored methods to incorporate sparsity in LLMs \cite{iclr_2024_relu, asplos_2022_dota, nips_2024_sparsellm} with comparable accuracy to dense counterparts. As depicted in Fig.~\ref{fig:llm_bkg}, the sparsity manifests in two forms: activation sparsity,  where intermediate results contain zeros, and weight sparsity, where model parameters contain zeros.

As Fig.~\ref{fig:llm_bkg} illustrates, sparsity exists across LLM layers, including $Q$, $K$, $V$, $O$ and $FC1$, $FC2$. However, sparsity levels vary greatly. For instance, activation sparsity in the $FC2$ can reach 97\%, while in the $FC1$ it typically ranges from 35\% to 70\%~\cite{iclr_2024_relu}. Moreover, application-specific accuracy requirements also affect the acceptable sparsity levels. This variability in sparsity patterns creates significant challenges for efficient tensor compression and data movement~\cite{isscm_2025_sparse}. To fully harness the benefits of sparsity, sparse accelerators need to adapt to these patterns and enable efficient LLM inference.

\subsection{Design Features of Sparse Accelerators} \label{subsec:accelerator_bkg}

\begin{figure}[t]
    \centering
    \includegraphics[width=0.48\textwidth]{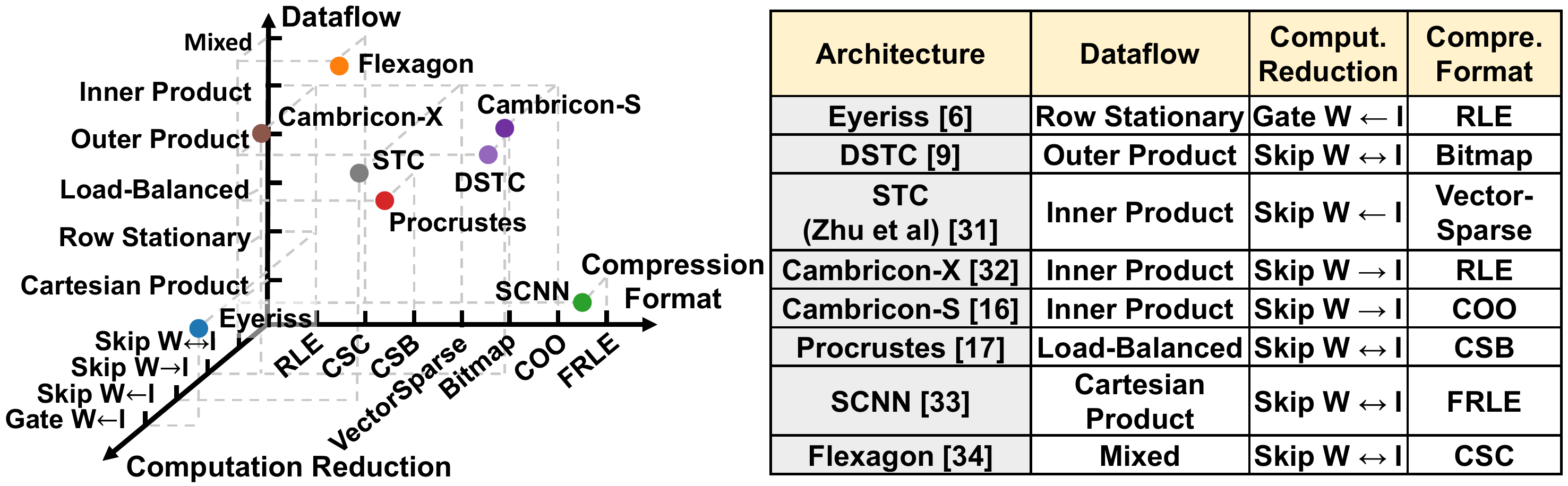}
    \caption{Sparse accelerators and their design features.}
    \label{fig:sparse_accelerators}
    \vspace{-0.5cm}
\end{figure}

Sparse accelerators~\cite{jssc_2017_eyeriss, micro_2019_stc, micro_2016_cambricon-x, micro_2018_cambricon-s, isca_2021_dstc, isca_2017_scnn, asplos_2023_flexagon} introduce additional sparsity mechanisms to better handle sparse models, offering higher speed and energy efficiency. As depicted in Fig.~\ref{fig:sparse_accelerators}, their features can be summarized in three aspects~\cite{micro_2022_sparseloop,tc_2021_zigzag}: dataflow, computation reduction and compression format.

\subsubsection{\textbf{Dataflow}}

Dataflow involves for-loop permutation combined with spatial and temporal mapping~\cite{tc_2021_zigzag}. Effective dataflow involves loop unrolling for parallelism, loop order optimization, and loop allocation to memory hierarchy to minimize energy and latency in memory transfers. In LLMs, the core operation, matrix multiplication (MatMul), can be expressed as $O[M][K] = \sum_N I[M][N] \times W[N][K]$, involving loops over $M$, $N$, and $K$ dimensions. Loops can be further split and reordered, known as loop ordering. Then the data associated with each loop is mapped to different levels of the memory hierarchy, which is called loop allocation. Sparse dataflow modeling has been extensively explored~\cite{micro_2022_sparseloop, song2024dimo}, providing a solid foundation for efficient accelerator design.

\subsubsection{\textbf{Computation Reduction}}
Sparse tensor operations improve performance by reducing unnecessary computations with gating and skipping. Gating idles MAC units on zero values, while skipping bypasses invalid operations. These mechanisms can be unidirectional (e.g., $\leftarrow$, $\rightarrow$) when checking a single operand, or bidirectional ($\leftrightarrow$) when checking both. For instance, Skipping W$\leftarrow$I executes computation only if the input is non-zero. With only five strategies and skipping typically performing best, this dimension requires little exploration.

\subsubsection{\textbf{Compression Format}}
Data compression is the key to optimizing sparse accelerator design by storing and transferring only non-zero elements, known as \textbf{payload}, along with their positions, or \textbf{metadata}. This reduces data movement across memory hierarchies, saving energy and lowering latency. Unlike dataflow and computation reduction, compression format optimization remains largely underexplored. Existing sparsity-aware DSE frameworks typically consider only limited, widely-used formats, lacking a systematic approach to design exploration. For example, DiMO-Sparse~\cite{song2024dimo} primarily focuses on data size after compression, neglecting important details like which tensor dimensions are compressed, significantly impacting data access overhead. Sparseloop~\cite{micro_2022_sparseloop} assumes uniform compression across all dimensions, leading to inaccuracies. More critically, current frameworks struggle to discover new formats due to reliance on a small set of predefined options. To bridge the critical gap, we propose SnipSnap, a DSE framework that systematically explores and optimizes compression formats for efficient sparse LLM inference.

\begin{figure}[t]
    \centering
    \includegraphics[width=0.49\textwidth]{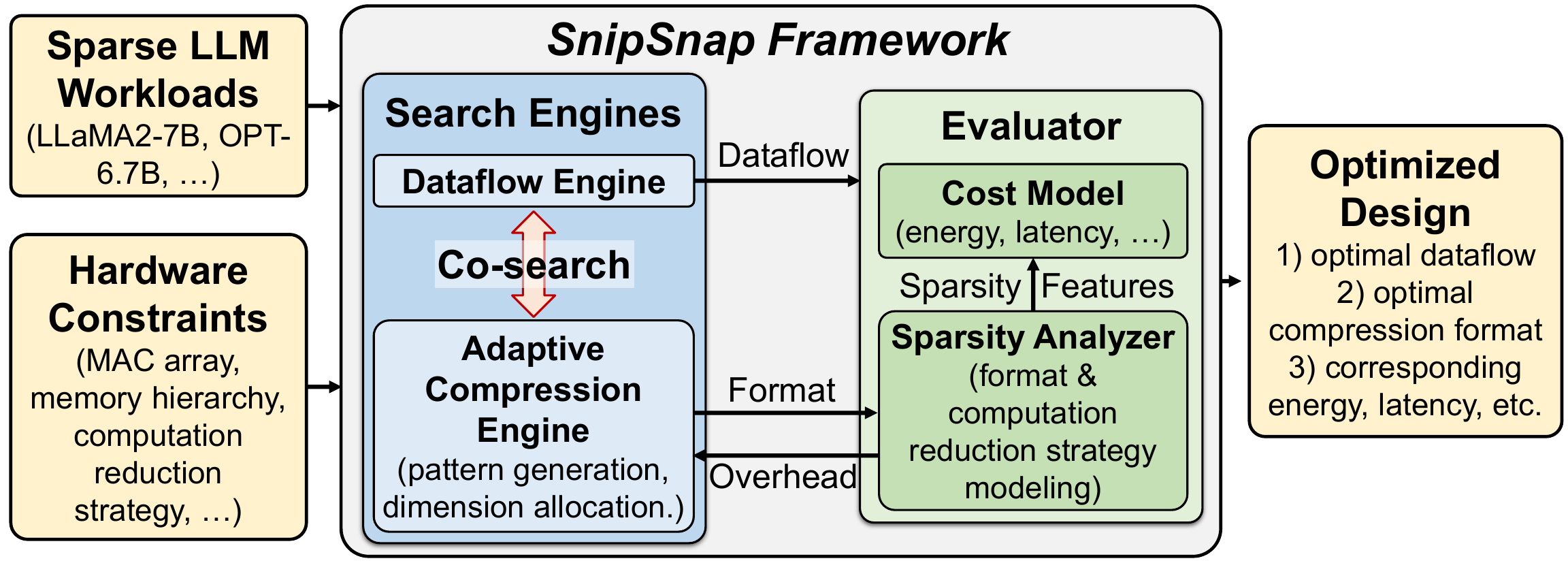}
    \vspace{-0.5cm}
    \caption{Overview of the proposed \projName~framework.}
    \label{fig:framework_diagram}
    \vspace{-0.5cm}
\end{figure}

\section{The Proposed Framework}
\label{sec:proposed_framwork}

\subsection{Overview}

As shown in Fig.~\ref{fig:framework_diagram}, \projName~automates the co-optimization of dataflow and compression format for sparse accelerators. It takes two inputs: (1) sparse workloads, possibly including one or multiple LLMs, with operator-level computation and sparsity specifications; and (2) hardware configurations, such as MAC array counts, memory hierarchy, and computation reduction strategies. Based on these, \projName~performs a two-phase optimization: the Search Engine generates design candidates, and the Evaluator selects the optimal design towards the prioritized performance metric.

The Search Engine includes two components.  The Dataflow Engine is responsible for identifying the optimal computation dataflow, including  the parallelism and scheduling strategies.
Since dataflow optimization has been extensively studied, \projName~adopts existing methodologies~\cite{tc_2021_zigzag,micro_2022_sparseloop}. To the best of our knowledge, the adaptive compression engine is proposed for the first time. It searches for effective compression formats and is detailed in Sec.~\ref{subsec:search_engine}.

The Evaluator is composed of a Sparsity Analyzer and a Cost Model. The Sparsity Analyzer estimates compressed data sizes and computation reduction using statistical expectations to quantify costs. The Cost Model evaluates hardware-specific metrics by modeling MAC operations and memory transfers. Together, they assess each design candidate's performance.

Finally, \projName outputs the optimal design configuration according to the specified priority target metric, including compression format, dataflow and corresponding performance metrics such as energy consumption, latency, and energy-delay-product (EDP).

\subsection{Hierarchical Compression Format Encoding} 
\label{subsec:compre_format}

We propose a hierarchical encoding method that models existing formats and structures the compression format space for efficient exploration. To manage its complexity, the space is divided into two subspaces: the \textbf{compression pattern space}, which defines the compression logic, and the \textbf{dimension allocation space}, which further specifies the allocation of compressed dimensions.

\begin{figure}[t]
    \centering    \includegraphics[width=0.48\textwidth]{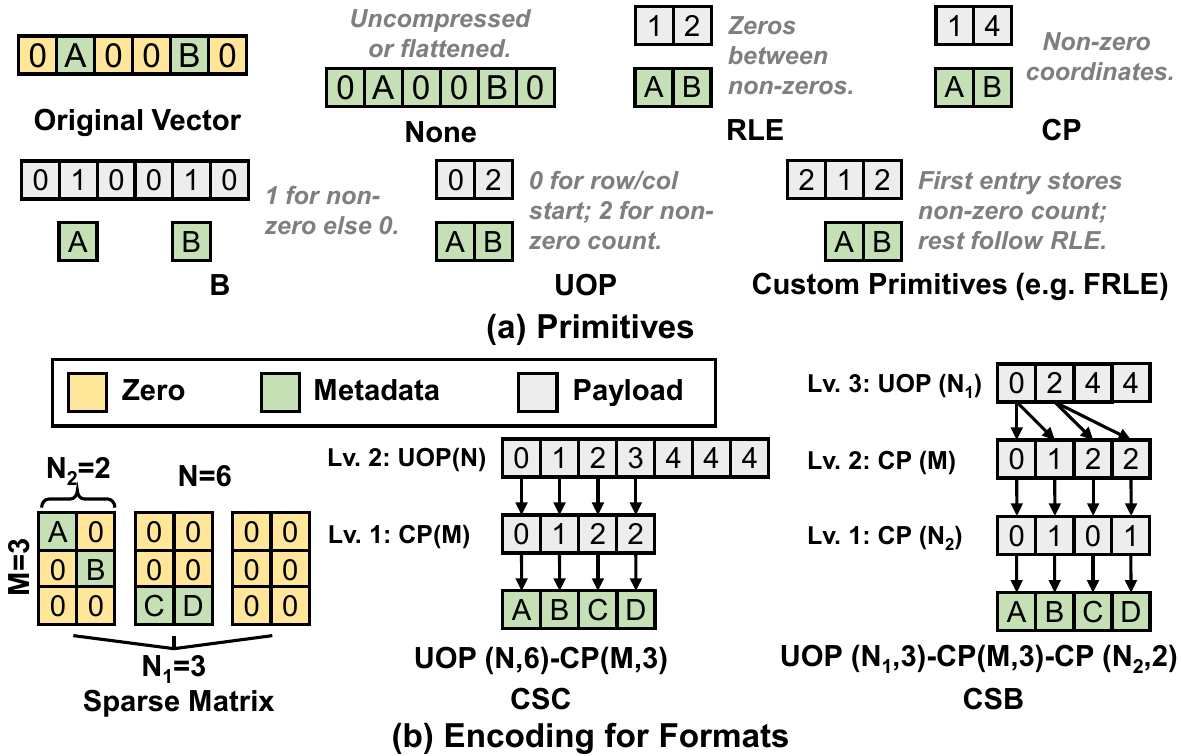}
    \caption{(a) Primitives for our hierarchical encoding method. (b) Examples of applying \texttt{CSB} (Procrustes~\cite{micro_2020_procrustes}), \texttt{CSC} (Flexagon~\cite{asplos_2023_flexagon}).}
    \label{fig:primitive}
    \vspace{-1em}
\end{figure}

\subsubsection{\textbf{Compression Pattern Space}}

A compression pattern is an ordered sequence of primitives, each treated as a level associated with specific dimensions and arranged from higher to lower levels.

\begin{definition}
\textbf{Compression Pattern Space ($CompPat$)}
\small
\[
CompPat(n) = [prim_1(dim_1), prim_2(dim_2), \dots, prim_n(dim_n)],
\]
\end{definition}
where $n\in \mathbb{Z}^+$ is the number of levels, $prim_i$ is the primitive at level $i$, and $dim_{i}$ denotes a dimension or subdimension.

We define primitives to represent basic compression operations, as shown in Fig.~\ref{fig:primitive}(a): \texttt{Run-Length Encoding} (\texttt{RLE}) represents the number of zeros between adjacent non-zeros. \texttt{Coordinate Payload} (\texttt{CP}) records the coordinates of non-zeros. \texttt{Bitmap} (\texttt{B}) uses a bit per element to indicate zero positions. \texttt{Uncompressed Offset Pairs} (\texttt{UOP}) \texttt{UOP} encodes group-wise first non-zero positions, ending with the total count. \texttt{None} indicates either an uncompressed or flattened dimension. \texttt{Custom} refers to user-defined primitives to capture operations not covered above. 

At this stage, we omit dimension sizes. For example, consider a \texttt{Compressed Sparse Column (CSC)} format over an $M \times N$ tensor is represented as \texttt{UOP($N$)-CP($M$)}, where \texttt{UOP} is the higher level.

\subsubsection{\textbf{Dimension Allocation Space}}
Once a compression pattern is defined, the final format is completed by assigning sizes to each subdimension. This derives from the prime factorization of the original dimension, and further optimization is discussed in Section~\ref{subsubsec:allocating}.
\begin{definition}
\textbf{Dimension Allocation Space ($DimAlloc$)}
\small
\[
DimAlloc(CompPat) = \{ (dim_{i,j}, size_{i,j}) \mid dim_{i,j} \in CompPat\},
\]
\end{definition}
where $CompPat$ is the compression pattern, $dim_{i,j}$ is the j-th subdimension of $dim_i$ and $size_{i,j}$ is the size of $dim_{i,j}$.

The compression pattern space and dimension allocation space together construct the full compression format exploration space. For example, a $M$$\times$$N$ tensor ($M$=3, $N$=6) with pattern \texttt{UOP($N_1$)-CP($M$)-CP($N_2$)} and $N$ split into subdimensions of size 3 and 2 results in \texttt{UOP($N_1$,3)-CP($M$,3)-CP($N_2$,2)}. Fig. \ref{fig:primitive}(b) further shows detailed encodings of \texttt{Compressed Sparse Block (CSB)} and \texttt{CSC}.

\begin{figure}[t]
    \centering    
    \includegraphics[width=0.48\textwidth]{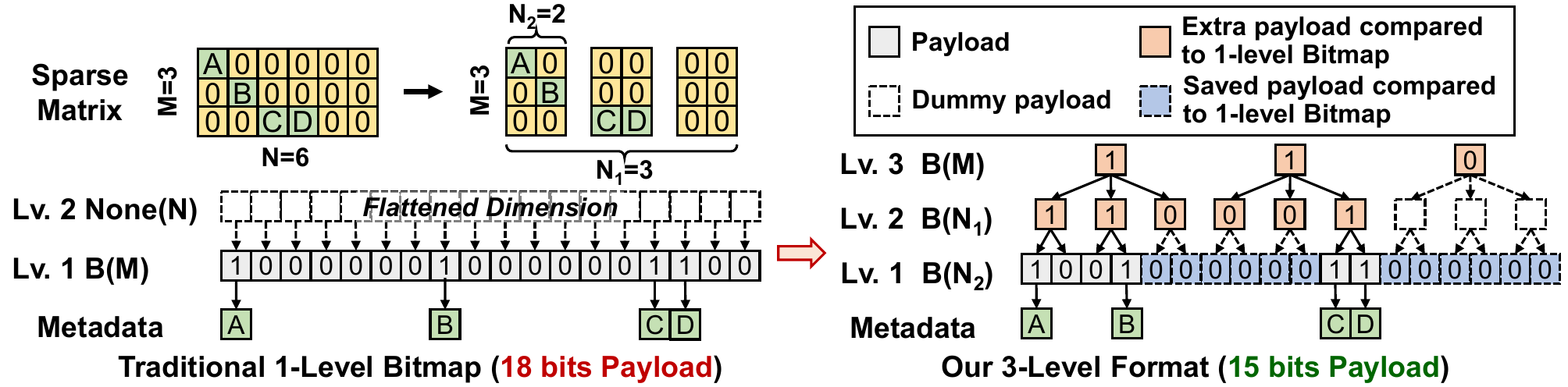}
    \caption{Comparison between the traditional one-level \texttt{B} format and a new three-level format exclusively enabled by our hierarchical representation method, achieving a 16.7\% reduction in payload.}
    \label{fig:fmt_comp}
    \vspace{-0.3cm}
\end{figure}

Our encoding enables discovering new, more efficient formats. For instance, as shown in Fig.~\ref{fig:fmt_comp}, using the \texttt{B} primitive, a matrix originally compressed in a single level can be encoded hierarchically across three levels. By splitting $N=6$, one bit can replace six, reducing the payload by 16.7\%.

\subsection{Adaptive Compression Engine}
\label{subsec:search_engine}

The adaptive compression engine generates candidate compression formats. To identify high-efficiency formats that generalize across tensors, we introduce three techniques:

\subsubsection{\textbf{Complexity-Based Penalizing}} 
While deeper compression patterns can reduce payload size (Fig.~\ref{fig:fmt_comp}), they often lead to over-specialized formats that are less efficient on other tensors with different sparsity. Moreover, more compression levels increases hardware complexity, which can undermine the advantages gained from compression. Hence, we introduce a penalty mechanism linked to the compression pattern level. We define the equivalent data size as $EqData = \gamma \times ActualData$, with a default $\gamma = 1.05^{\text{level}}$ that balances accuracy and remains configurable. During pattern search, formats with higher $EqData$ than simpler ones are excluded, as their payload savings do not justify the loss in generality.

\begin{figure}[t]
    \centering
    \includegraphics[width=0.46\textwidth]{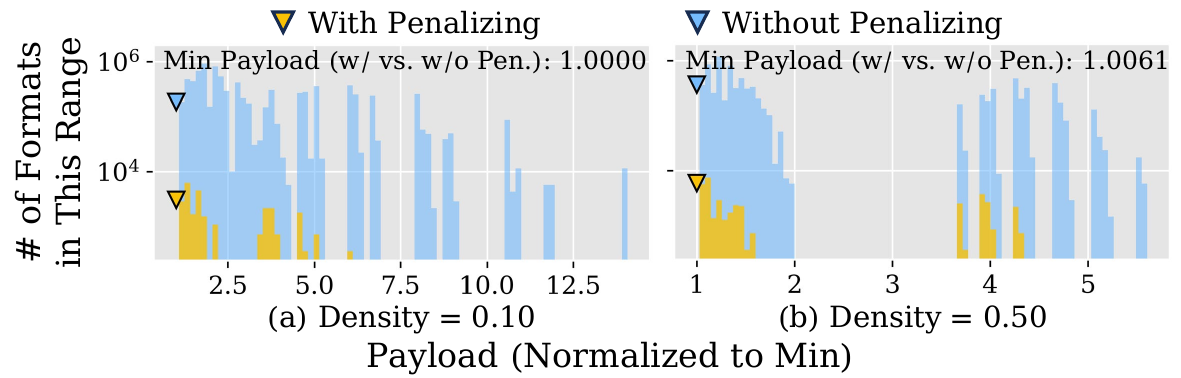}
    \caption{Identified total payloads of different formats with and without complexity-based penalizing. Complexity-based penalizing efficiently finds near-optimal formats while significantly reducing compression complexity.}
    \label{fig:compression_payload}
    \vspace{-0.5cm}
\end{figure}

As shown in Fig.~\ref{fig:compression_payload}, when applying this penalty in the search over a $4096\times4096$ matrix, with 90\% and 2:4 sparsity, respectively, it significantly reduces the number of explored patterns from over 400{,}000 to a much smaller subset, while still maintaining a near-optimal payload size within 0.31\% and improving generality with typically only 2 to 3 levels.

\subsubsection{\textbf{Efficiency-Oriented Allocating}}
\label{subsubsec:allocating}
For a given compression pattern, the variety of subdimension decompositions leads to multiple possible dimension allocations, each affecting the cost of compression and decompression. A good allocation should simplify this process, reducing related overhead to improve overall efficiency. To achieve this, we prioritize subdimension decomposing and allocation based on the dataflow's loop ordering configuration. 
For example, given a format \texttt{B($M_1$)-B($M_2$)} and a loop ordering where $M$=8 is the outer loop and $M$=32 is the inner loop, setting $M_1$=8 and $M_2$=32 aligns the compression format with the dataflow, reducing runtime overhead~\cite{asplos_2023_flexagon}. Although other combinations such as (32, 8) or (64, 4) are possible, they may lead to misalignment and increased cost.

\subsubsection{\textbf{Importance-Based Scoring}} 
In real deployments, multiple LLMs with different structures, sparsity levels, and usage frequencies may share the same sparse accelerator. Designing a universally efficient format is challenging. To address this, we introduce an importance score ($ImpScore$) to weight operators based on usage frequency or performance priority. For instance, if OPT-125M handles most tasks while OPT-6.7B is used rarely (1\% of cases), their scores can be set to 99 and 1, respectively. Based on previous techniques, we first identify the optimal format per LLM and then select the final format by minimizing the weighted optimization metric ($OptMetric$):
$$\underset{format}{\operatorname{argmin}}\sum_{i} ImpScore(LLM_i) \times OptMetric(LLM_i).$$

Furthermore, \projName accommodates advanced scenarios such as accelerators supporting multiple compression formats~\cite{icce_2024_conversion}. For these cases, it adapts by disabling importance-based scoring to prevent format conflicts.

\subsection{Progressive Co-Search Workflow} 
\label{subsec:co-consideration}

\begin{figure}[t]
    \centering
    \includegraphics[width=0.46\textwidth]{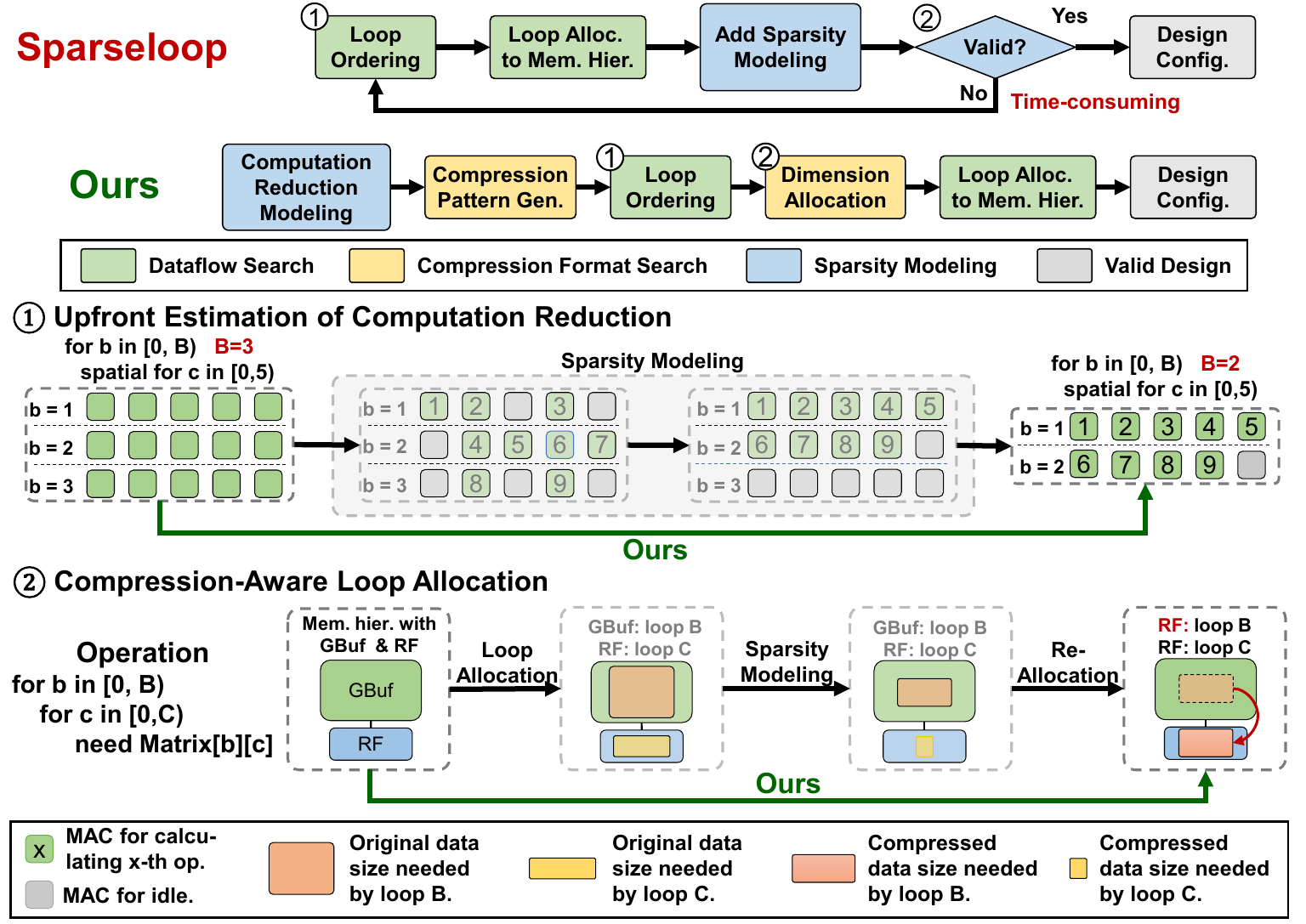}
    \caption{Workflow comparison between Sparseloop and \projName.}
    \label{fig:co-consideration}
    \vspace{-0.5cm}
\end{figure}

Given the large design space of dataflow and compression formats, an efficient workflow is crucial for identifying Pareto-optimal designs within a reasonable time. However, current sparsity-aware DSE frameworks, such as Sparseloop~\cite{micro_2022_sparseloop}, suffer from inefficiencies caused by stepwise modeling. Its workflow is illustrated in Fig.~\ref{fig:co-consideration}, where the green and blue shade refer to dataflow search and sparse feature modeling. Sparseloop first searches dataflow for dense workloads, then modifies configurations to account for sparsity features such as compression and computation reduction, followed by legality checks. This iterative correction process leads to redundant modeling and slow exploration.

As depicted in Fig.~\ref{fig:co-consideration}, \projName addresses these limitations through a progressive co-search workflow that interleaves  dataflow and compression format exploration. The workflow begins with the Sparsity Analyzer modeling the computation reduction strategy. Based on this, \projName generates compression patterns, determines loop ordering for each pattern and refines formats through dimension allocation. It then maps for-loops to the memory hierarchy based on the compressed data size estimated by the Sparsity Analyzer. This progressive co-search workflow is supported by two key techniques:

\subsubsection{\textbf{Upfront Estimation of Computation Reduction}}
\projName eliminates post-hoc corrections by estimating sparse operations before dataflow generation. For example, skipping reduces actual computations and shrinks temporal loop bounds. By incorporating this upfront, \projName captures the impact of computation reduction strategy from the outset.

\subsubsection{\textbf{Compression-aware Loop Allocation}}
Compression alters tensor sizes, often requiring loop reallocation in traditional workflows. \projName avoids this by modeling compressed sizes, perform dimension allocation to determine post-compression shapes before allocating for-loops. This ensures generated dataflow remains valid without later adjustments.

\section{Experimental Results} \label{sec:exp}

\subsection{Experiment Setup} 

\subsubsection{\textbf{Hardware Configurations}} 
The evaluation is conducted on various sparse accelerators~\cite{jssc_2017_eyeriss, isca_2017_scnn, isca_2021_dstc}. All configurations are summarized in Table~\ref{table:arch_config}. To support LLM inference, all accelerators are scaled to 16$\times$ MACs and 4$\times$ on-chip memory.

\subsubsection{\textbf{LLM Benchmarks and Compression Format Baselines}} We employ various sparse LLMs from~\cite{iclr_2024_relu, nips_2024_sparsellm} with multiple sizes, including LLaMA2~\cite{arxiv_2023_llama2}(7B, 13B) and OPT~\cite{arxiv_2022_opt}(6.7B-30B). For format comparisons, we select four widely-used formats as baselines~\cite{arxive_2021_sparsity}: \texttt{Bitmap}, \texttt{RLE}, \texttt{CSR}, and \texttt{COO}.

\subsubsection{\textbf{Evaluation Methodologies}}
\textbf{(a) Modeling accuracy validation (Sec.~\ref{subsec:acc_val}).} We validate \projName's energy and latency estimation using SCNN\cite{isca_2017_scnn} and DSTC~\cite{isca_2021_dstc}, comparing with published data. Mean relative error is used to quantify accuracy.
\textbf{(b) Compression format identification (Sec.~\ref{subsec:fmt_perf}).} We evaluate our adaptive compression engine through memory energy consumption and speedup under various sparsity levels. Activation and weight sparsity are evaluated separately to show their distinct impacts. We choose the state-of-the-art (SotA) \textbf{Arch 3} as the primary accelerator~\cite{isca_2021_dstc} for these evaluations.
For single-model evaluation, we compare against the best of the four standard baselines. We further extend our evaluation to multi-model inference scenarios, adding baselines with per-model optimal formats to evaluate generalization.
\textbf{(c) Exploration speed assessment (Sec.~\ref{subsec:speed_validation}).}
We evaluate our co-search methodology by comparing \projName with Sparseloop~\cite{micro_2022_sparseloop} and DiMO-Sparse~\cite{song2024dimo}. For Sparseloop, we measure energy efficiency under a 20-minute per-MatMul time limit across 10 runs, reporting both search time and solution quality. We test \projName in both fixed-format and search-enabled modes. We further estimate \projName's speedup qualitatively. 
\textbf{(d) Format Feasibility Discussion (Sec.~\ref{subsec:discuss}).} We further present optimized formats from \projName and analyze their hardware overhead to validate deployment feasibility.

\begin{figure}[t]
    \centering\includegraphics[width=0.46\textwidth]{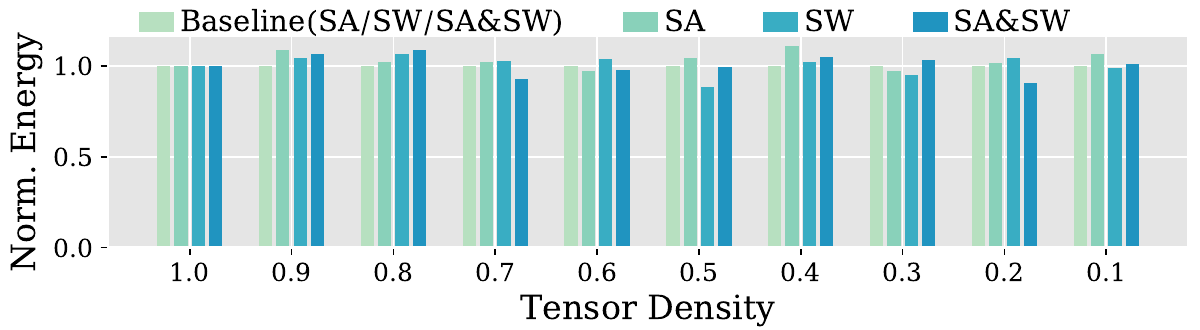}
    \vspace{-1em}
    \caption{Energy validation for \projName. Results for sparse activations (SA), sparse weights (SW), and both (SA\&SW) are normalized to reported data~\cite{isca_2017_scnn}.}
    \label{fig:scnn_energy}
    \vspace{-1em}
\end{figure}

\begin{figure}[t]
    \centering\includegraphics[width=0.46\textwidth]{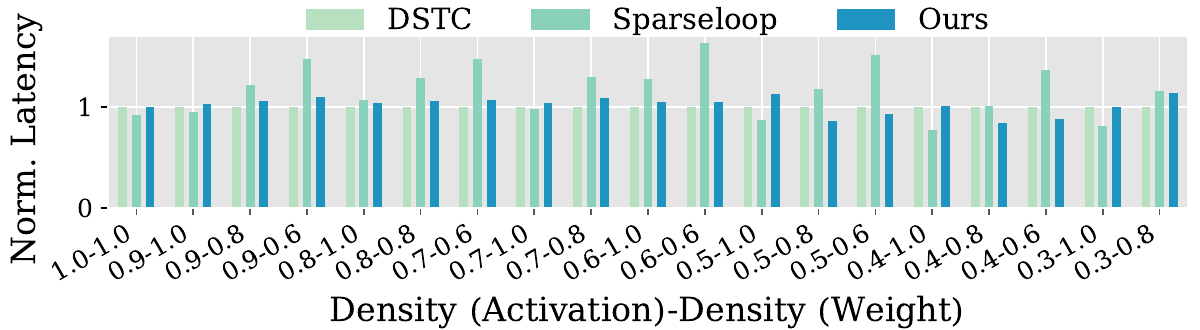}
    \vspace{-1em}
    \caption{Latency validation for \projName on DSTC~\cite{isca_2021_dstc}.}
    \label{fig:dstc_latency}
    \vspace{-1em}
\end{figure}

\begin{figure*}[t]
    \centering    
    \includegraphics[width=0.98\textwidth]{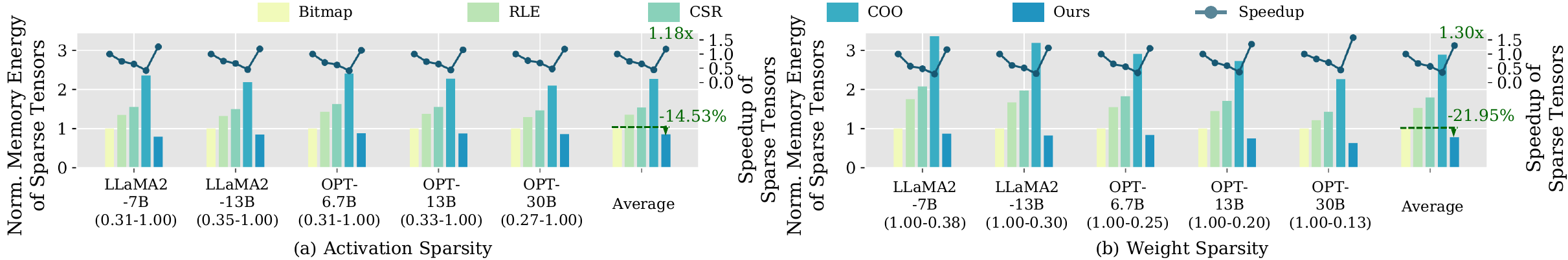}
    \vspace{-0.6em}
    \caption{Memory energy consumption and speedup of various sparse LLMs using different compression formats. Results are normalized to Bitmap.}
    \vspace{-1em}
    \label{fig:fmt_perf_single}
\end{figure*}

\begin{figure}[t]
    \centering    
    \includegraphics[width=0.48\textwidth]{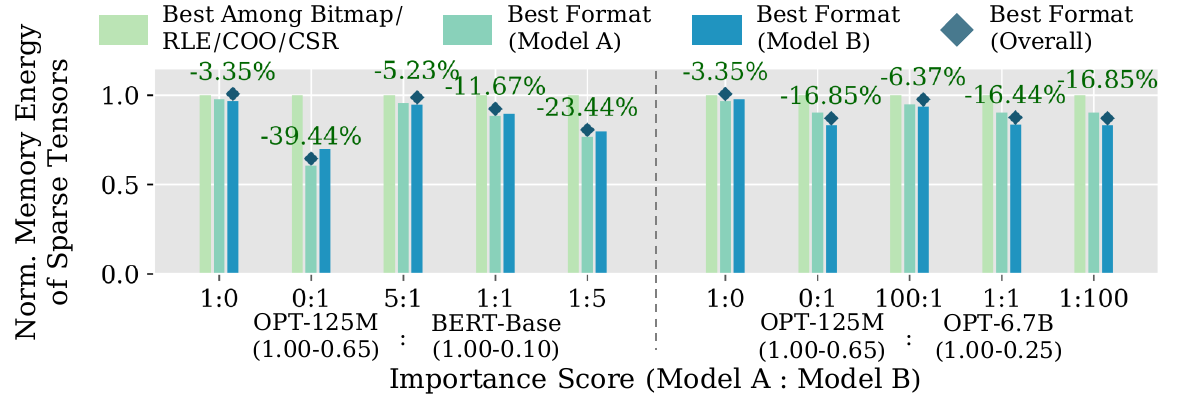}
    \vspace{-0.6em}
    \caption{Normalized energy consumption of multiple sparse LLMs (BERT-Base+OPT-125M and OPT-125M+OPT-6.7B) with varying importance score.}
    \label{fig:fmt_perf_multi}
    \vspace{-0.5em}
\end{figure}

\subsection{Modeling Accuracy Validation} \label{subsec:acc_val}

For energy, we choose SCNN~\cite{isca_2017_scnn}, modeling its architecture and comparing results with published data. As shown in Fig.~\ref{fig:scnn_energy}, \projName achieves a mean relative error of 4.33\% across sparse activation (SA), sparse weight (SW), and combined (SA\&SW) cases. For latency, we compare \projName with DSTC's reported results and Sparseloop~\cite{micro_2022_sparseloop}, using its open-source artifact. The test focuses on 4096$\times$4096 MatMul under various sparsity levels common in LLaMA2-7B~\cite{arxiv_2023_llama2}. As shown in Fig.~\ref{fig:dstc_latency}, \projName achieves an average error of 6.26\%, outperforming Sparseloop’s 8.55\%.

\begin{table*}[t]
    \centering
    \caption{\projName's Actual Modeling Time (in seconds) and Its Exploration Speedups (in \textcolor{SpdClr}{Green}) over the Current SotA Sparseloop}
    \label{table:speed_validation}
    \resizebox{0.97\textwidth}{!}{
    \begin{tabular}{ccccccccccc} 
    \toprule
    \cmidrule(lr){2-11}
    \multirow{2}{*}{Arch.} & \multicolumn{2}{c}{LLaMA2-7B} & \multicolumn{2}{c}{LLaMA2-13B} & \multicolumn{2}{c}{OPT-6.7B} & \multicolumn{2}{c}{OPT-13B} & \multicolumn{2}{c}{OPT-30B}\\
    \cmidrule(lr){2-11}
    &  Fixed & Search & Fixed & Search & Fixed & Search & Fixed & Search & Fixed & Search \\
    \midrule
    Arch 1 & 37.0$^{\textcolor{SpdClr}{2300.6\times}}$ & 472.4$^{\textcolor{SpdClr}{180.3\times}}$ & 44.4$^{\textcolor{SpdClr}{2351.2\times}}$ & 671.2$^{\textcolor{SpdClr}{155.6\times}}$ & 34.1$^{\textcolor{SpdClr}{2464.6\times}}$ & 352.2$^{\textcolor{SpdClr}{238.5\times}}$ & 44.9$^{\textcolor{SpdClr}{2297.4\times}}$ & 453.7$^{\textcolor{SpdClr}{227.5\times}}$ & 65.2$^{\textcolor{SpdClr}{2172.4\times}}$ & 578.3$^{\textcolor{SpdClr}{244.9\times}}$ \\
    Arch 2 & 22.6$^{\textcolor{SpdClr}{3769.2\times}}$ & 198.9$^{\textcolor{SpdClr}{428.4\times}}$ & 23.7$^{\textcolor{SpdClr}{4406.1\times}}$ & 217.8$^{\textcolor{SpdClr}{479.2\times}}$ & 20.7$^{\textcolor{SpdClr}{4060.3\times}}$ & 267.4$^{\textcolor{SpdClr}{314.2\times}}$ & 21.5$^{\textcolor{SpdClr}{4804.6\times}}$ & 216.4$^{\textcolor{SpdClr}{476.9\times}}$ & 40.3$^{\textcolor{SpdClr}{3517.4\times}}$ & 314.7$^{\textcolor{SpdClr}{450.0\times}}$ \\
    Arch 3 & 32.3$^{\textcolor{SpdClr}{2635.4\times}}$ & 306.5$^{\textcolor{SpdClr}{277.9\times}}$ & 41.7$^{\textcolor{SpdClr}{2502.8\times}}$ & 370.2$^{\textcolor{SpdClr}{282.0\times}}$ & 32.2$^{\textcolor{SpdClr}{504.2\times}}$ & 300.5$^{\textcolor{SpdClr}{54.1\times}}$ & 39.3$^{\textcolor{SpdClr}{476.6\times}}$ & 369.4$^{\textcolor{SpdClr}{50.7\times}}$ & 55.7$^{\textcolor{SpdClr}{435.8\times}}$ & 508.8$^{\textcolor{SpdClr}{47.7\times}}$ \\
    Arch 4 & 28.8$^{\textcolor{SpdClr}{2957.3\times}}$ & 301.7$^{\textcolor{SpdClr}{282.4\times}}$ & 48.8$^{\textcolor{SpdClr}{2138.1\times}}$ & 371.5$^{\textcolor{SpdClr}{281.0\times}}$ & 40.3$^{\textcolor{SpdClr}{404.5\times}}$ & 297.4$^{\textcolor{SpdClr}{54.9\times}}$ & 47.0$^{\textcolor{SpdClr}{415.1\times}}$ & 365.6$^{\textcolor{SpdClr}{53.4\times}}$ & 68.1$^{\textcolor{SpdClr}{352.5\times}}$ & 502.0$^{\textcolor{SpdClr}{47.8\times}}$ \\
    \bottomrule
    \end{tabular}
    }
    \vspace{-1.2em}
\end{table*}

\subsection{Compression Format Optimization Results} \label{subsec:fmt_perf}

We conduct two experiments to present the effectiveness of our adaptive compression engine. The first targets varying sparsity within a single LLM, while the second assesses its ability to balance compression across multiple LLMs.

In the first experiment, we evaluate sparse LLMs from~\cite{iclr_2024_relu, nips_2024_sparsellm} with 2048-token prefill and 128-token decoding~\cite{isca_2024_llmcompass}. Fig.~\ref{fig:fmt_perf_single} presents memory energy consumption and speedups normalized to \texttt{Bitmap}, with activation-weight density pairs indicated for each model. Among baselines, \texttt{Bitmap} performs best at typical LLM sparsity levels, while \texttt{CSR} and \texttt{COO} better suit highly sparse tensors. \projName outperforms the best baseline \texttt{Bitmap}, achieving average 14.53\% energy savings with 1.18$\times$ speedup for activation sparsity, and 21.95\% savings with 1.30$\times$ speedup for weight sparsity. Larger models, with higher sparsity, benefit more from our multi-level compression.

The second experiment evaluates two multi-model scenarios. Case one runs BERT-Base~\cite{arxiv_2018_bert} processing 256 input tokens for natural language understanding and OPT-125M~\cite{arxiv_2022_opt} processing 256 input tokens and generating 32 output tokens for text generation. Case two adopts a speculative decoding setup~\cite{leviathan2023fast} with OPT-125M and OPT-6.7B~\cite{arxiv_2022_opt}, both processing 256 input tokens and generating 32 output tokens.

Fig.~\ref{fig:fmt_perf_multi} shows energy results, normalized to the best of baseline formats. \projName consistently outperforms them, averaging 14.23\% energy savings. This is enabled by importance-based scoring, which biases format selection toward higher importance. In case one, emphasizing BERT-Base boosts savings due to its higher sparsity; in case two, format design should prioritize OPT-6.7B due to its significant higher cost.

\subsection{Exploration Speed Results}
\label{subsec:speed_validation}

To evaluate \projName's exploration efficiency, we first compare it with Sparseloop. Table~\ref{table:speed_validation} demonstrates \projName's  runtime in seconds with speedup over Sparseloop indicated in green superscripts. Both activation and weight densities are set to 0.75. ``Fixed" uses preset formats, while ``Search" enables format exploration. Architectures are detailed in Table~\ref{table:arch_config}.

As shown in Table~\ref{table:speed_validation}, \projName achieves average speedups of 2248.3$\times$ (Fixed) and 231.46$\times$ (Search) over Sparseloop. This efficiency comes from our progressive co-search workflow, which uses upfront computation estimation and compression-aware loop allocation to avoid stepwise modeling overhead. Although enabling format search adds some cost, it is justified by the performance gains detailed in Sec.~\ref{subsec:fmt_perf}. Compared to Sparseloop, \projName explores both dataflow and compression formats far more efficiently, enabling practical design optimization for LLM accelerators.


We further compare with DiMO-Sparse~\cite{song2024dimo} with preset compression formats. While DiMO-Sparse is limited to CNNs, \projName generalizes across workloads and is evaluated on AlexNet~\cite{nips_2012_alexnet}, VGG-16~\cite{corr_2014_vgg} and ResNet-18~\cite{arxiv_2015_resnet}, achieving 19.4$\times$, 19.7$\times$ and 23.8$\times$speedups, highlighting both \projName’s scalability and substantial efficiency gains.

\begin{table}[t]
    \centering
    \caption{Hardware Configurations of Evaluation}
    \resizebox{0.48\textwidth}{!}{
        \begin{tabular}{ccccc}
        \toprule
             & Arch 1 & Arch 2 & Arch 3 & Arch 4 \\
        \midrule
        MACs & 2688 & 2688 & 2048 & 2048 \\
        Mem. Hierarchy & Eyeriss~\cite{jssc_2017_eyeriss} & Eyeriss~\cite{jssc_2017_eyeriss} & DSTC~\cite{isca_2021_dstc} & DSTC~\cite{isca_2021_dstc} \\ 
        \makecell{Comp. Reduc.\\Strategy} & Gating I$\rightarrow$W & Skipping I$\rightarrow$W & Skipping I$\leftrightarrow$W & Gating I$\leftrightarrow$W \\
        Compress. format & RLE & RLE & Bitmap & Bitmap \\
        \bottomrule
        \end{tabular}
    }
    \label{table:arch_config}
\end{table}

\subsection{Format Feasibility Discussion}
\label{subsec:discuss}

To illustrate format selection in Sec.~\ref{subsec:fmt_perf}, we showcase two examples:
For weight-sparse OPT-6.7B in Fig.~\ref{fig:fmt_perf_single}, SnipSnap selects
\texttt{B($M$)-B($N$)-B($N$)} for $M$$\times$$N$ tensors, the same format shown in Fig.~\ref{fig:fmt_comp}. For BERT-Base in Fig~\ref{fig:fmt_perf_multi}, \texttt{UOP($M$)-B($N$)} replaces \texttt{CSR}’s \texttt{CP} with a lower-overhead \texttt{B}, improving efficiency at that sparsity level.

With only 2$\sim$3 levels enabled by complexity-based penalizing, our formats show the simplicity similar to \texttt{CSR} and \texttt{CSB} and incur similarly low hardware cost. Existing sparse accelerators report compression/decompression area overheads of 1.56\%–15.45\%~\cite{isca_2021_dstc,isca_2017_scnn,hpca_2020_sigma, micro_2020_procrustes, hpca_2022_s2ta, jsscc_2020_snap}, which is justified by the significant performance gains from effective compression, as shown in Sec.~\ref{subsec:fmt_perf}.

\section{Conclusion} \label{sec:conclusion}

This paper presents \projName, a compression format and dataflow co-optimization framework for efficient sparse LLM accelerator design. It introduces: (1) a hierarchical format encoding that broadens the design space; (2) an adaptive compression engine that selects formats for tensors with varying sparsity, achieving 18.24\% average memory energy savings; and (3) a progressive co-search workflow that jointly optimizes dataflow and compression, delivering average speedups of 2248.3$\times$ over Sparseloop and 21.0$\times$ over DiMO-Sparse.

\bibliographystyle{IEEEtran}

\bibliography{ref}

\end{document}